# Image quality assessment for closed-loop computer-assisted lung ultrasound

Zachary M C Baum[a,b], Ester Bonmati[a,b], Lorenzo Cristoni[c], Andrew Walden[d], Ferran Prados[a],
Baris Kanber[a], Dean C Barratt[a,b], David J Hawkes[a,b], Geoffrey J M Parker[a,e],
Claudia A M Gandini Wheeler-Kingshott[e,f*], Yipeng Hu[a,b*]

[a]Centre for Medical Image Computing, University College London;
[b]Wellcome / EPSRC Centre for Surgical and Interventional Sciences, University College London;
[c]Frimley Park Hospital, Frimley Health NHS Foundation Trust;
[d]Royal Berkshire Hospital, Royal Berkshire NHS Foundation Trust;
[e]NMR Research Unit, Queen Square MS Centre, UCL Queen Square Institute of Neurology;
[f]Department of Brain and Behavioural Sciences, University of Pavia

*Equal contribution as senior authors*

## ABSTRACT

We describe a novel, two-stage computer assistance system for lung anomaly detection using ultrasound imaging in the intensive care setting to improve operator performance and patient stratification during coronavirus pandemics. The proposed system consists of two deep-learning-based models: a quality assessment module that automates predictions of image quality, and a diagnosis assistance module that determines the likelihood-of-anomaly in ultrasound images of sufficient quality. Our two-stage strategy uses a novelty detection algorithm to address the lack of control cases available for training the quality assessment classifier. The diagnosis assistance module can then be trained with data that are deemed of sufficient quality, guaranteed by the closed-loop feedback mechanism from the quality assessment module. Using more than 25,000 ultrasound images from 37 COVID-19-positive patients scanned at two hospitals, plus 12 control cases, this study demonstrates the feasibility of using the proposed machine learning approach. We report an accuracy of 86% when classifying between sufficient and insufficient quality images by the quality assessment module. For data of sufficient quality – as determined by the quality assessment module – the mean classification accuracy, sensitivity, and specificity in detecting COVID-19-positive cases were 0.95, 0.91, and 0.97, respectively, across five holdout test data sets unseen during the training of any networks within the proposed system. Overall, the integration of the two modules yields accurate, fast, and practical acquisition guidance and diagnostic assistance for patients with suspected respiratory conditions at point-of-care.

## 1. INTRODUCTION

Lung ultrasound (LUS) imaging can be used for point-of-care diagnosis and management of some pulmonary conditions in patients presented with respiratory symptoms [1], such as COVID-19-related pneumonia [2]. When outbreaks of infectious respiratory diseases happen, bedside LUS imaging at point-of-care offers unique accessibility and flexibility for healthcare provision while helping to manage infection control [3]. Indeed, studies have indicated that LUS may be useful for assessing suspicious COVID-19-positive cases for triage during the early management of the disease [4]. As such; rapid triage with LUS could be potentially useful even when swab tests using polymerase chain reaction (PCR) or lateral flow tests are available. A prospective study on 41 patients showed that LUS not only detects positive COVID-19 patients but also indicates prognosis, evidenced by the association with admission to intensive care units and death [5]. As LUS sees potential use in rapidly-evolving pandemic situations, it stands as a means of use in future major epidemics. The sensitivity of LUS in diagnosis and prognostication of the COVID-19 cases also has potential for a much-needed monitoring tool for short-term hospitalized patients and long-term recovery, although the prognosis and monitoring roles of LUS are not discussed further in this work.

Despite positive results from various referral centers, and the widespread availability of US equipment, the use of LUS for assisting with the management of infections such as COVID-19 is severely limited, particularly in smaller hospitals and countries with limited healthcare infrastructure. We argue that this lack of adoption is likely to be due partly to the lack of healthcare professionals with the requisite training to acquire and correctly interpret diagnostic-quality US images, and the practical challenges associated with providing this training [6]. Modern machine learning methods, often represented by deep learning, can potentially address these problems through automated LUS image classification. Born *et al*. [7] reported an overall 89% and 92% classification accuracy between three classes of lung ultrasound images (COVID-19, bacterial pneumonia, and healthy) using neural networks, at the frame and video level, respectively. However, Born *et al.*'s method reports a false positive rate of 0.55 in healthy images, revealing that many healthy images were incorrectly classified as containing COVID-19 features by the neural network. Furthermore, Roy *et al.* [8] proposed deep learning models for frame- and video-level severity score prediction, as well as pixel-level segmentation, with promising clinical utility if the accuracy is translated to day-to-day intensive care units or emergency departments.

However, such methods are often trained and tested using expert-acquired US data. In addition to the feasibility in obtaining large quantities of such data for training, the use of high-quality training images may mandate that the same requirement of image quality be reliably maintained during the inference stage when these machine learning models are used in clinical practice. It is widely acknowledged that there is a steep learning curve associated with high-quality US data acquisition, and substantial variation has been observed in acquisition and interpretation protocols between experts [9]. Therefore, image quality assessment should be at the centre of developing any machine learning models that are intended to be deployed clinically to inform patient care.

We propose a computer-assisted system for both US acquisition and interpretation, which aids clinical diagnosis and prognosis through the prediction of both image quality and the risk of anomaly. In particular, we present a closed-loop design of an imaging acquisition protocol based on a novel image quality assessment algorithm. This design aims to ensure downstream image interpretation tasks receive data of sufficient quality, such that the accuracy of the clinical use of LUS, here detection of anomaly, can be safeguarded or even improved. We also propose a one-class novelty detection network to address the use of non-representative training data that have few or limited negative examples, which is a common issue in the context of a rapidly evolving epidemics, such as at the start of the SARS-COV-2 pandemic [10]. This paper presents the application of the proposed algorithms to detect COVID-19-positive patients, demonstrating the feasibility of using a quality assessment module in conjunction with a diagnosis assistance module. We conducted experiments using a set of retrospective LUS data from both a control group and a positive COVID-19 group, confirmed by swab tests using PCR.

Our contributions include: 1) a closed-loop imaging acquisition assistance system based on image quality assessment as a feedback mechanism; 2) a novel image quality assessment module that combines supervised classification with one-class novelty detection networks; and 3) a rigorous validation experiment using COVID-19 LUS data obtained on patients at the peak of the pandemic.

## 2. METHODS

### 2.1 Closed-loop lung ultrasound system

Figure 1 provides an overview of our proposed system. The quality assessment module classifies US images and provides user feedback. When image quality is insufficient, users are instructed to repeat the acquisition at this location on the patient, following application-specific protocols. Although considered out of scope for this work, such a system may prompt users to repeat predefined protocols, for example making use of external assistance such as teleguidance or automated action suggestions [11, 12]. The diagnosis assistance module uses images determined to be of sufficient quality to perform predefined tasks, such as inferring the likelihood of specific conditions.

### 2.2 Quality assessment module

We consider two strategies for training convolutional neural networks to accept or reject diagnostic-quality LUS images: The first strategy uses a classification network that receives a LUS frame and predicts class probabilities for individual label types – sufficient and insufficient quality. A binary classification network ($QA^{bin}$) based on the Visual Geometry Group (VGG) network architecture [13] is used to discriminate images of sufficient quality from those of insufficient quality. Training $QA^{bin}$ requires manual labeling of the data as belonging to one of the two classes, sufficient quality or insufficient quality. The second strategy is to train an adversarial deep learning model for one-class novelty detection [14], which requires only examples of images of sufficient quality. A common scenario encountered during a rapidly-evolving epidemic is that control data from healthy subjects or patients with other conditions, obtained using the same protocol, are not readily available [10]. A quality assessment novelty detection ($QA^{nd}$) approach is proposed to train a reconstructor network and a discriminator network in an adversarial manner [14]. Using this approach, the reconstructor network learns how to reconstruct the target class – images with sufficient quality in this case – while it produces larger errors when reconstructing images outside of the target distribution, assessed by the simultaneously-trained discriminative network.

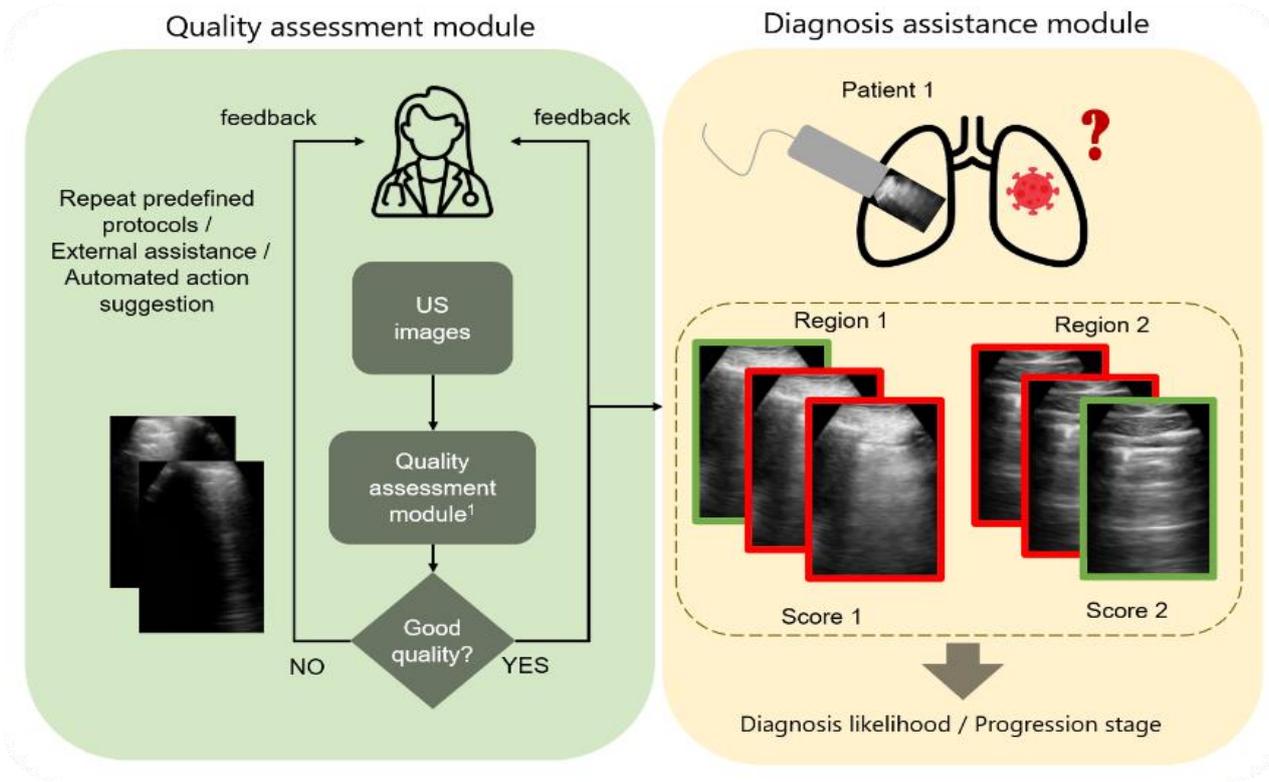

Figure 1. Illustration of the proposed closed-loop lung ultrasound system described in Section 2.

Additionally, we propose combining these two models using Bayesian model averaging. This third quality assessment method ($QA^{bin+nd}$) assesses image quality by assessing the weighted quality scores: $p_{QA}(x) = p^{bin}p_{bin}(x) + p^{nd}p_{nd}(x)$, where $p_{bin}(x)$ and $p_{nd}(x)$ are class probabilities obtained from $QA^{bin}$ and $QA^{nd}$, respectively. The prior probabilities, $p^{bin}$ and $p^{nd}$, are estimated by the proportions of respective training data sizes. $QA^{bin+nd}$ serves as a simple mechanism to combine the two previous models, which may be trained with different data sets acquired at different stages of the epidemic. As such, model re-training may not be required as new data becomes available or if there is restricted access to data. The proposed averaging method, including the one-class classifier, potentially maximizes the use of available data for training while reducing the potential bias due to varying data representativeness in this application [10].

### 2.3 Diagnosis assistance module

A second binary classification network ($D^{bin}$) is trained, also based on the VGG network [13], to predict whether the given LUS image frames are in the COVID-19-positive- or the COVID-19-negative class. The use of well-established network architectures, such as the VGG network used in $D^{bin}$, allows this work to investigate the feasibility of the feedback mechanism enabled by the quality assessment module.

The overall classification errors of $D^{bin}$ on independent test data sets, sensitivity and specificity at various threshold values are reported. These results are further compared with those calculated using only images that were previously classified by the quality assessment module as being of sufficient quality. However, further work is needed to test this strategy with video-level or subject-level classification tasks and for more fine-grained severity scores or prognostic clinical outcomes.

### 2.4 Data

Patient images were obtained in two hospitals by two clinicians, both using a Butterfly iQ US probe (Butterfly Inc., Guilford, CT, USA). In total, 25800 LUS images were acquired at 10 frames per second from 37 COVID-19-positive patients (11495 images from 10 positive cases at Site A and 14305 images from 27 positive patients at Site B). Additionally, 16627 images from 12 control patients, who tested negative for COVID-19, were acquired at Site A. All COVID-19 diagnoses were confirmed by PCR tests. Image quality was manually labeled as sufficient or insufficient by an experienced ultrasound imaging researcher with over 5 years of experience with clinical US imaging. This resulted in 699 images from Site A and 238 images from Site B being denoted as of insufficient quality. Example images are provided in Figure 2.

**2.5 Experiments**

Due to challenges with control data availability across imaging sites, this work adopts a two-way split for training and test sets without the use of a validation set. This prohibits network fine-tuning through methods such as hyperparameter tuning. The proposed quality assessment networks, $QA^{bin}$, $QA^{nd}$, and $QA^{bin+nd}$, were trained on data from Site B, which includes only COVID-19-positive patient images. $QA^{bin}$ was trained on both sufficient- and insufficient-quality images, whilst $QA^{nd}$ was trained on images of sufficient quality. The diagnosis assistance module network $D^{bin}$ was trained five times, with a five-fold cross-validation, using Site A data that includes both positive and control images. The Site A data were partitioned at patient-level into five sets of training and test sets.

Three experiments are presented to evaluate the three methods for constructing the quality assessment module, $QA^{bin}$, $QA^{nd}$, and $QA^{bin+nd}$, by assessing the performance on the diagnosis assistance module network $D^{bin}$. The five Site A test sets (from the five-fold validation) were used as this data was not previously seen by any network training and contained both positive and control images. Before classification for diagnosis assistance, each fold of data was "screened" independently by each quality assessment network, yielding three "screened" versions of each Site A test set. Each version is then classified by $D^{bin}$. Testing in this manner provides three sets of diagnosis assistance module results on three different "screened" versions of the Site A test sets. This facilitates validation of the entire proposed system using independent test data (i.e. data unseen during the training of the networks used for testing quality assessment or diagnosis assistance).

All neural networks were implemented in TensorFlow 2 [15] and Keras [16]. Reference-quality open-source code was adopted with default hyperparameter parameter and training configurations (see [13] and [14]), where possible, for reproducibility. The quality assessment and diagnosis assistance modules were trained on an NVIDIA DGX-1 using a Tesla V100 GPU for approximately 90 and 60 minutes each, respectively.

## 3. RESULTS

We present qualitative examples of classifications from both quality assessment (Figure 2) and diagnosis assistance modules (Figure 3). Guided gradient-weighted class activation maps [17] were computed to color-code the activated regions during the classification predictions, with red to blue indicating decreasing activation. This visualization, shown in Figure 3, provides potential regions of anatomical and pathological interest, suggesting that the networks may have utilized interesting, meaningful and human interpretable LUS features, which may be effectively used for diagnosis assistance and further improvement of the diagnosis assistance system. However, future research is required to quantitatively investigate the interpretation of the class activation maps and their clinical values.

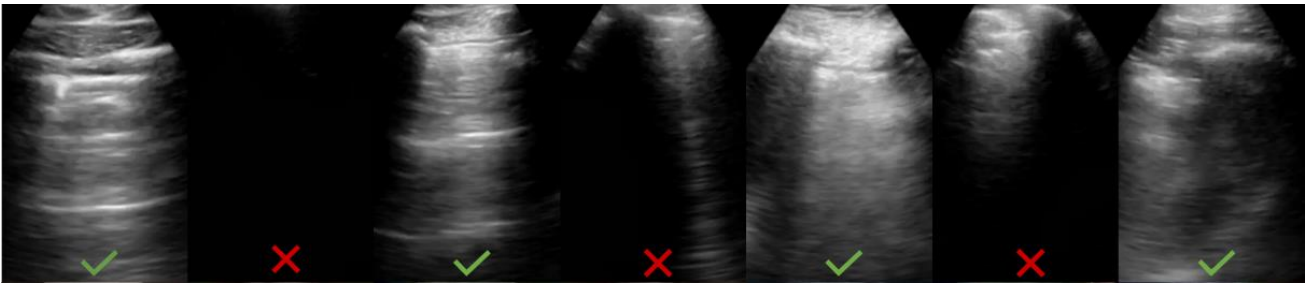

Figure 2. Example LUS images from the quality assessment module, green ticks & red crosses indicating classified as being of sufficient and insufficient quality.

Table 1 summarizes the experimental results described in Section 2.5. The overall classification accuracy (the rate of correct classification) of the proposed quality assessment module is 0.85 using $QA^{bin}$ or $QA^{nd}$ alone, and 0.86 using the combined $QA^{bin+nd}$. The classification accuracy for the diagnosis assistance module, $D^{bin}$, without any quality assessment is 0.95, with specificity reaching 1.00, 0.98, and 0.98, when sensitivity is 0.80, 0.90, and 0.95, respectively. Classification accuracy of the diagnosis assistance module, $D^{bin}$, following quality assessment, and rejection of images of insufficient quality, is 0.95, 0.97, and 0.95 when using $QA^{bin}$, $QA^{nd}$, and $QA^{bin+nd}$, respectively.

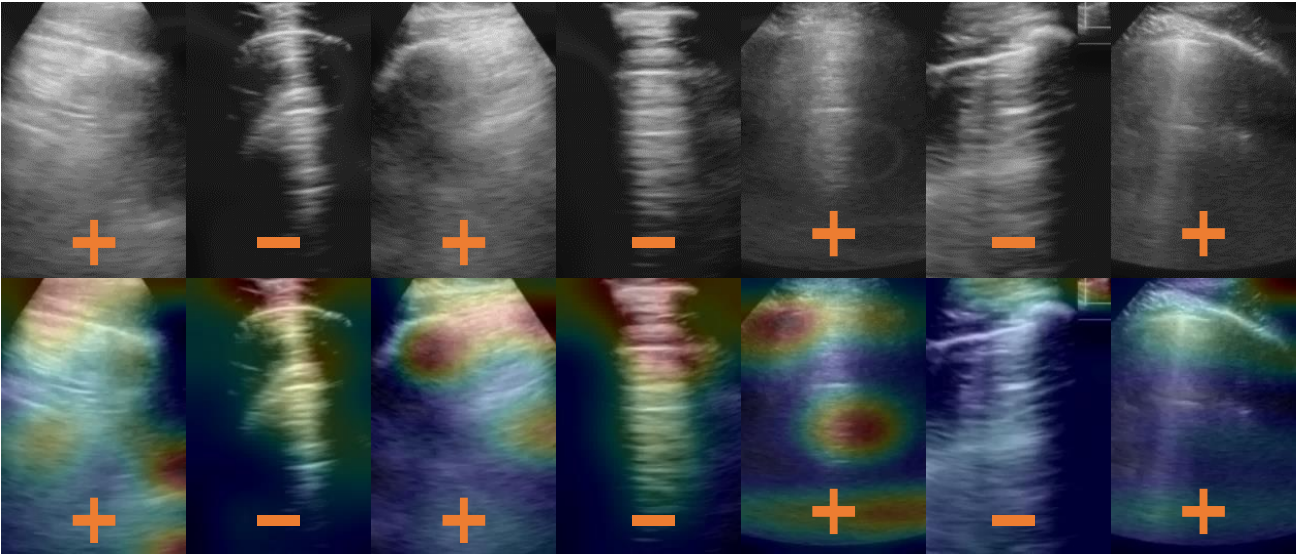

Figure 3. Example LUS images from $D^{bin}$; top row: + and – signs indicating true-positive and true-negative COVID-19 diagnosis, bottom row: the same images overlaid with guided gradient-weighted class activation maps.

It is important to note that a decrease in classification accuracy to 0.86 was observed when testing on data of insufficient quality. The employed dataset has many fewer negative examples for training and testing, an over-optimistic scenario compared to what would be expected in real-life clinical LUS procedures. Nevertheless, the improvements seen when trained with more data of insufficient quality are likely to be greater and more impactful for less experienced users. Similar constraints on the diagnosis assistance module need to be considered when interpreting the results reported in this work. The experimental data were acquired by experienced clinicians, although there is still a significant proportion of frames that decreased performance of the diagnostic accuracy without prior quality assessment.

Table 1. Summary of the classification accuracy, sensitivity and specificity at various threshold values.

| Method | Quality assessment | | Diagnosis assistance with $D^{bin}$ | | |
|---|---|---|---|---|---|
| | Classification Accuracy | TP, TN, FP, FN | Classification Accuracy | Sensitivity w.r.t. Specificity=0.8, 0.9, 0.95 | Specificity w.r.t. Sensitivity=0.8, 0.9, 0.95 |
| $QA^{bin}$ | 0.856 | 0.853, 0.991, 0.009, 0.147 | 0.950 | 0.999, 0.996, 0.970 | 0.996, 0.983, 0.979 |
| $QA^{nd}$ | 0.851 | 0.850, 0.889, 0.111, 0.150 | 0.965 | 0.997, 0.970, 0.957 | 0.987, 0.967, 0.961 |
| $QA^{bin+nd}$ | 0.861 | 0.858, 0.995, 0.005, 0.142 | 0.952 | 0.999, 0.995, 0.970 | 0.997, 0.983, 0.979 |

## 4. CONCLUSION

This work is the first, to our knowledge, to provide machine learning assistance for the acquisition and diagnostic classification using point-of-care LUS, validated on clinically acquired data from COVID-19-positive patients. Our system addresses an often-overlooked issue in LUS, ensuring images are of sufficient quality before machine learning assisted diagnosis. We have shown the feasibility of using LUS for diagnosis assistance in detecting positive COVID-19 cases from healthy controls. Our novel method combines a fully supervised classifier with a novelty detection algorithm to form a quality assessment module, which is validated with a diagnosis assistance module, used to predict abnormalities in LUS images. It is perhaps important to point out that the use of PCR test results as reference standard inevitably generated false negative labels. However, the approach to improve the reference standard, such as by combining with other clinically relevant, yet also imperfect, outcome measures, depends on use scenario of the system. For these specific clinical applications of the proposed system, such as maintaining high sensitivity or minimizing false negative cases, the quality assessment module could play an arguably more crucial role by altering the threshold to meet the desirable accuracy requirement during ultrasound acquisition. Moreover, the system has the potential to be extended to other clinical tasks for diagnosis, staging, and predicting clinical outcomes during LUS procedures, and will be validated prospectively by providing feedback to inexperienced clinicians.

## ACKNOWLEDGEMENTS

Z.M.C. Baum is supported by the Natural Sciences and Engineering Research Council of Canada Postgraduate Scholarships-Doctoral Program, the University College London Overseas and Graduate Research Scholarships. B. Kanber and F. Prados are supported by the NIHR Biomedical Research Centre at University College London Hospitals NHS Foundation Trust and University College London. G.J.M. Parker receives funding from EPSRC (EP/S031510/1). C.A.M. Gandini Wheeler-Kingshott is supported by the MS Society (#77), Wings for Life (#169111), Horizon2020 (CDS-QUAMRI, #634541), BRC (#BRC704/CAP/CGW), and the UCL Global Challenges Research Fund (GCRF). This work is also supported by the Wellcome/EPSRC Centre for Interventional and Surgical Sciences (203145Z/16/Z). G.J.M. Parker is a Director and Shareholder in Bioxydyn Limited. G.J.M. Parker and C.A.M. Gandini Wheeler-Kingshott are shareholders in Queen Square Analytics Limited.